\pdfoutput=1
\documentclass{easychair}

\usepackage{subfigure}

\newcommand{\xf}[1]{Figure~\ref{#1}}

\newcommand{\xs}[1]{Section~\ref{#1}}

\newcommand{\cpp}{{C++\index{C++}}}

\newcommand{\file}[1]{\url{#1}\index{Files!#1}}

\newcommand{\api}[1]{\texttt{#1}\index{API!#1}}

\newcommand{\macos}[1]{\index{Mac OS #1@{\sc{Mac OS #1}}}{\sc{Mac OS #1}}}

\newcommand{\win}[1]{\index{Windows #1@{\sc{Windows #1}}}{\sc{Windows #1}}}

\newcommand{\lucidL}[1]{{$\mathit{Lucid}$}($L$) }

		{}

\def\myvert{\raise 2.27pt \hbox{\vrule depth 0pt height 8pt width 0.2mm}}
\def\myarrow{\hspace*{0.43mm}%
             \raise 2.29pt\hbox{\vrule depth 0pt height 8pt width 0.16mm}%
             \hspace*{-0.32mm}%
             $\longrightarrow$
             \ %
             }

\newcommand{\oglsf}{OGLSF\index{Frameworks!OGLSF}\index{OpenGL Slides Framework}}

\newcommand{\slideimagewidth}{\columnwidth}

\begin{document}

\title{Teaching Physical Based Animation\\via OpenGL Slides}
\titlerunning{Teaching Physical Based Animation via OpenGL Slides}

\author{
Miao Song\\
Graduate School,\\
Concordia University,\\Montreal, Canada,\\
Email: \url{m_song@cse.concordia.ca}\\
\and
Serguei A. Mokhov\\
Computer Science\\and Software Engineering,\\
Concordia University, Montreal, Canada,\\
Email: \url{mokhov@cse.concordia.ca}\\
\and
Peter Grogono\\
Computer Science\\and Software Engineering,\\
Concordia University, Montreal, Canada,\\
Email: \url{grogono@cse.concordia.ca}
}

\authorrunning{Song, Mokhov, and Grogono}

\maketitle

\begin{abstract}
This work expands further our earlier poster presentation
and integration of the OpenGL Slides Framework ({\oglsf}) -- to make presentations
with real-time animated graphics where each slide is a scene with tidgets -- and physical based
animation of elastic two-, three-layer softbody objects. The whole project is very interactive,
and serves dual purpose -- delivering the teaching material in a classroom setting
with real running animated examples as well as releasing the source code
to the students to show how the actual working things are made.
\end{abstract}

\section{Introduction}
\label{sect:introduction}
It is very helpful for effective teaching of computer graphics (CG)
techniques~\cite{gtt,cgems,teaching-graphics-with-osl}, especially
advanced topics such as real-time physical based animation
of softbody objects, with a real working code on hand that can
be demonstrated in a classroom and then given out to students
for learning purposes and extension for their course work.
It is reasonable to assume, in subjects like CG, teaching may
be less effective if the examples are not visualized in
class for the students.
On top of that, it can be a nuisance 
for the instructor presenting the concepts and switching
between the presentation power-point-like slides
and the demo especially if it is complex and highly interactive
with a lot of variable parameters to tweak.
As a result, for the cases like the one briefly
described, we argue that it is more effective {\em and} efficient
to combine the OpenGL CG programs with OpenGL presentation
slides in one teaching unit. There the traditional power-points and various
techniques can be exemplified at run-time at the same time and the source
code can be released to the students later to follow the
examples through at all angles.
For this purpose we integrated the physical-based softbody simulation
system~\cite{msong-mcthesis-2007,softbody-framework-c3s2e08,softbody-lod-glui-cisse08}
with the OpenGL slides presentation framework ({\oglsf})~\cite{opengl-slides-cisse08,opengl-slides-grapp09}
that compose in a small demo that we discuss throughout this work.

\subsection*{Organization}
In \xs{sect:related work} we discuss the background and the related
work done that contribute to the creation of this teaching unit,
specifically we discuss the properties of the {\oglsf} in \xs{sect:oglsf}
and the Softbody Simulation System in \xs{sect:softbody-framework-system}.
We then describe a brief methodology and layout in \xs{sect:methodology}.
Afterwards, we conclude in \xs{sect:conclusion} describing our achievement,
the limitation of the approach in \xs{sect:limitations}, and the
future work items in \xs{sect:future-work}.
All the sections are illustrated with the actual screenshots from the
said OpenGL softbody system presentation slides and referenced where
appropriate.

\section{Related Work}
\index{related work}
\label{sect:related work}
The major pieces of the related work that contribute
to this works, are two frameworks alongside with their
implementation, put together with other items,
to eventually form a teaching module for computer graphics.
Here we extrapolate from our previous poster presentation~\cite{softbody-opengl-slides}
on this topic with more details on the actual design and 
implementation of the physical-based softbody
animation techniques in an OpenGL power-point-like
presentation tool with the demonstrated results. We describe the two CG systems
in this section in some detail for the unaware reader.

\subsection{OpenGL Slides Framework (OGLSF)}
\label{sect:oglsf}

\begin{figure*}[htp!]
\hrule\vskip4pt
\begin{center}
	\subfigure[Welcome Title Slide]
	{\label{fig:welcome}
	 \includegraphics[width=.47\slideimagewidth]{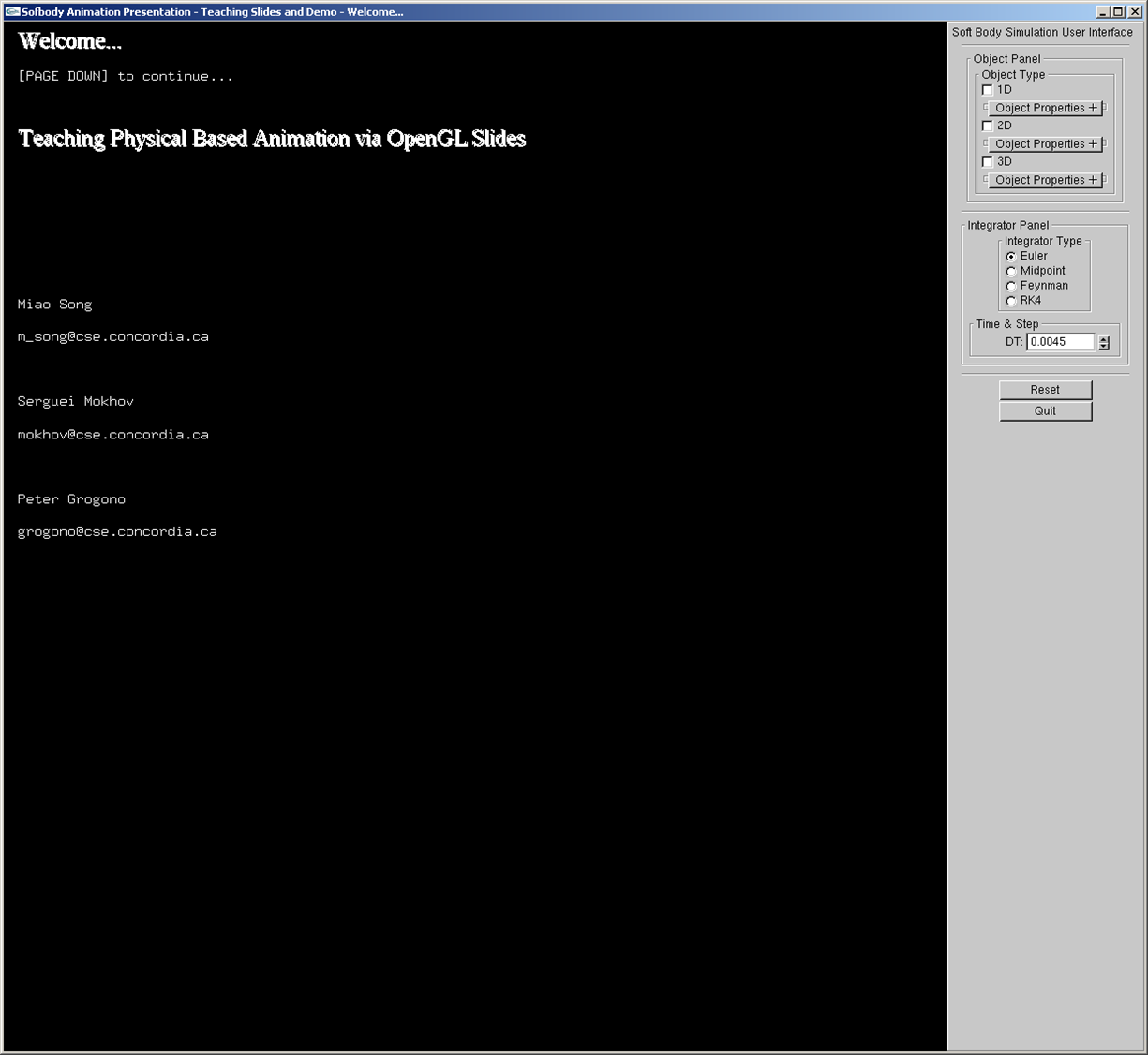}}
	\subfigure[Table of Contents Slide]
  {\label{fig:contents}
	 \includegraphics[width=.47\slideimagewidth]{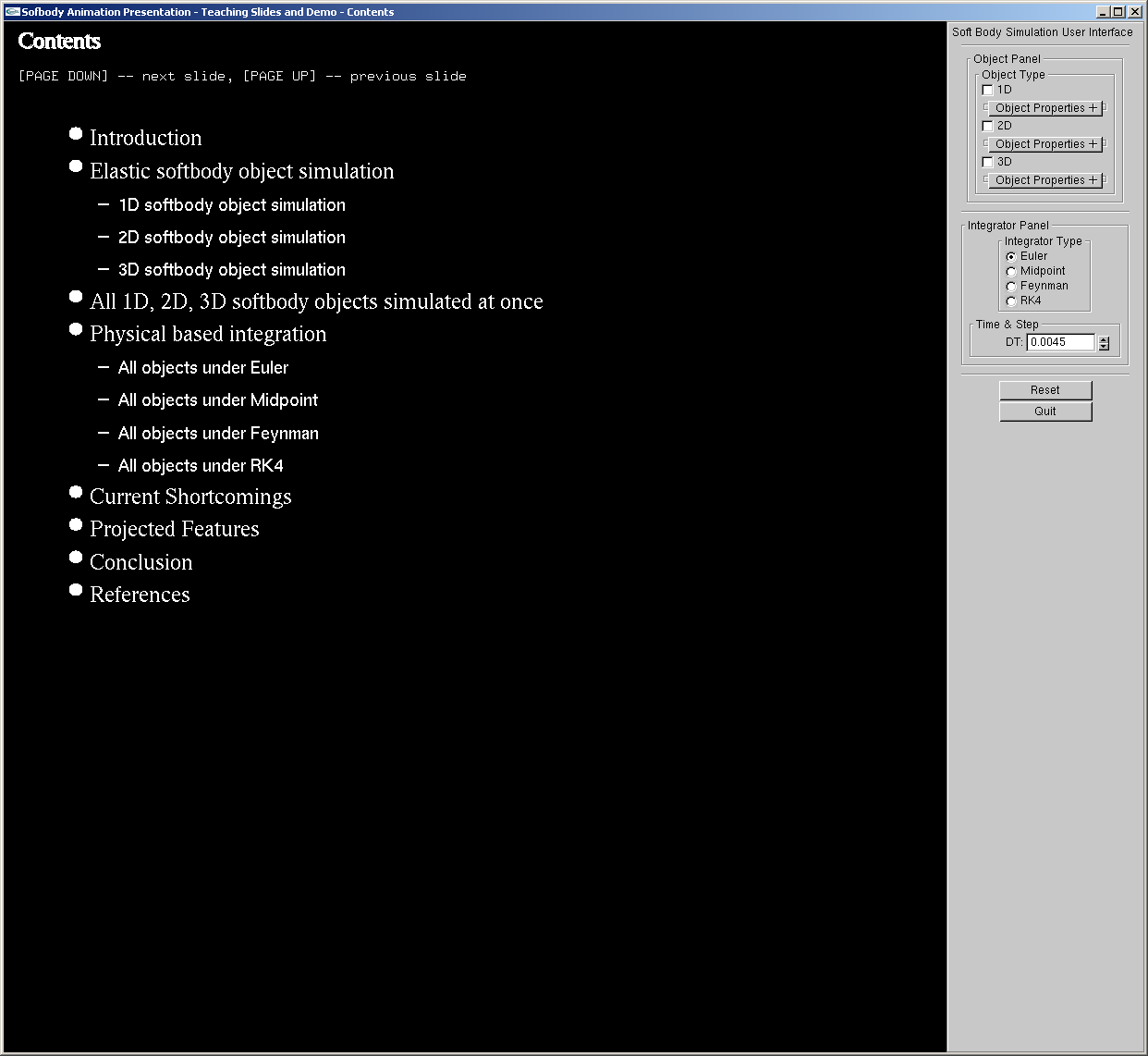}}
	\subfigure[Introduction Slide]
  {\label{fig:introduction}
	 \includegraphics[width=.47\slideimagewidth]{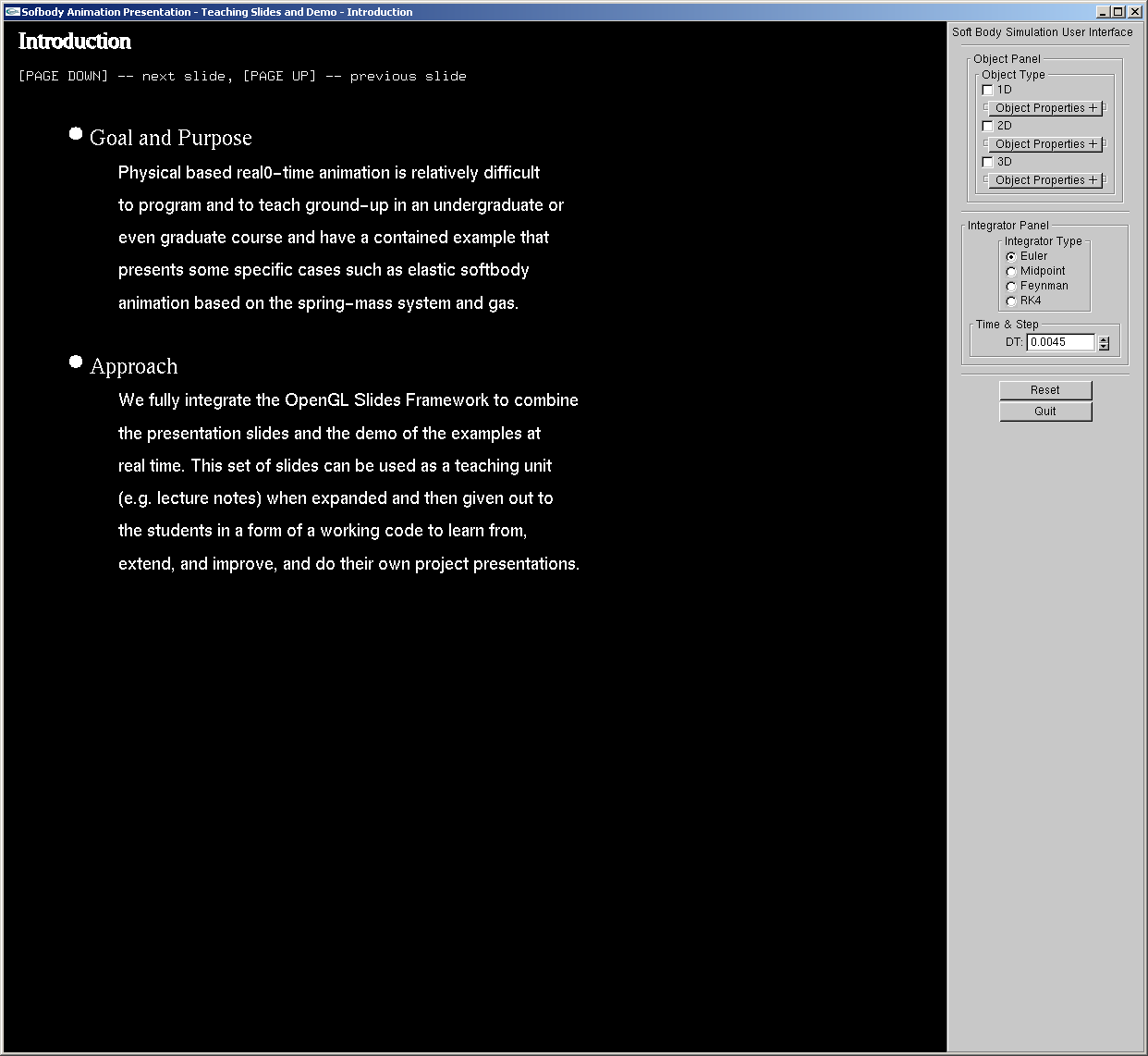}}
\caption{A Set of Introductory Tidget-only Slides}
\label{fig:introductory-slides}
\end{center}
\hrule\vskip4pt
\end{figure*}

{\oglsf} gives ability to make slides, navigate between them
using various controls, and allow for common bulleted textual widgets -- the {\em tidgets}.
It also allows to override the control handling from the main
idle loop down to each individual (current) slide.
All slides together compose a concrete instance of \api{Presentation},
which is a collection of slides
that uses the Builder pattern to sequence the slides. Each
slide is a derivative of the generic \api{Slide} class and represents
a scene with the default keyboard controls for the tidgets
and navigation.
It is understood
that the tidgets can be enabled and disabled to allow the
main animation to run unobstructed~\cite{opengl-slides-cisse08,opengl-slides-grapp09,softbody-opengl-slides}.
Each scene on the slide is modeled using traditional
procedural modeling techniques~\cite{wiki:procedural-modeling} and is set as a developer
or artist desires. It can include models and rendering
of any primitives, complex scenes, texturing, lighting,
GPU-based shading, and others, as needed and is fit
by the presenter~\cite{opengl-slides-cisse08,opengl-slides-grapp09}.
The main program delegates its handling of the callback
controls for keyboard, mouse, and idle all the way down to the
presentation object that handles it and passes it down
to each current slide~\cite{opengl-slides-cisse08,opengl-slides-grapp09,softbody-opengl-slides}.

\subsection{Softbody Simulation Framework and System}
\label{sect:softbody-framework-system}

\begin{figure*}[htp!]
	\hrule\vskip4pt
	\begin{center}
	\includegraphics[width=\textwidth]{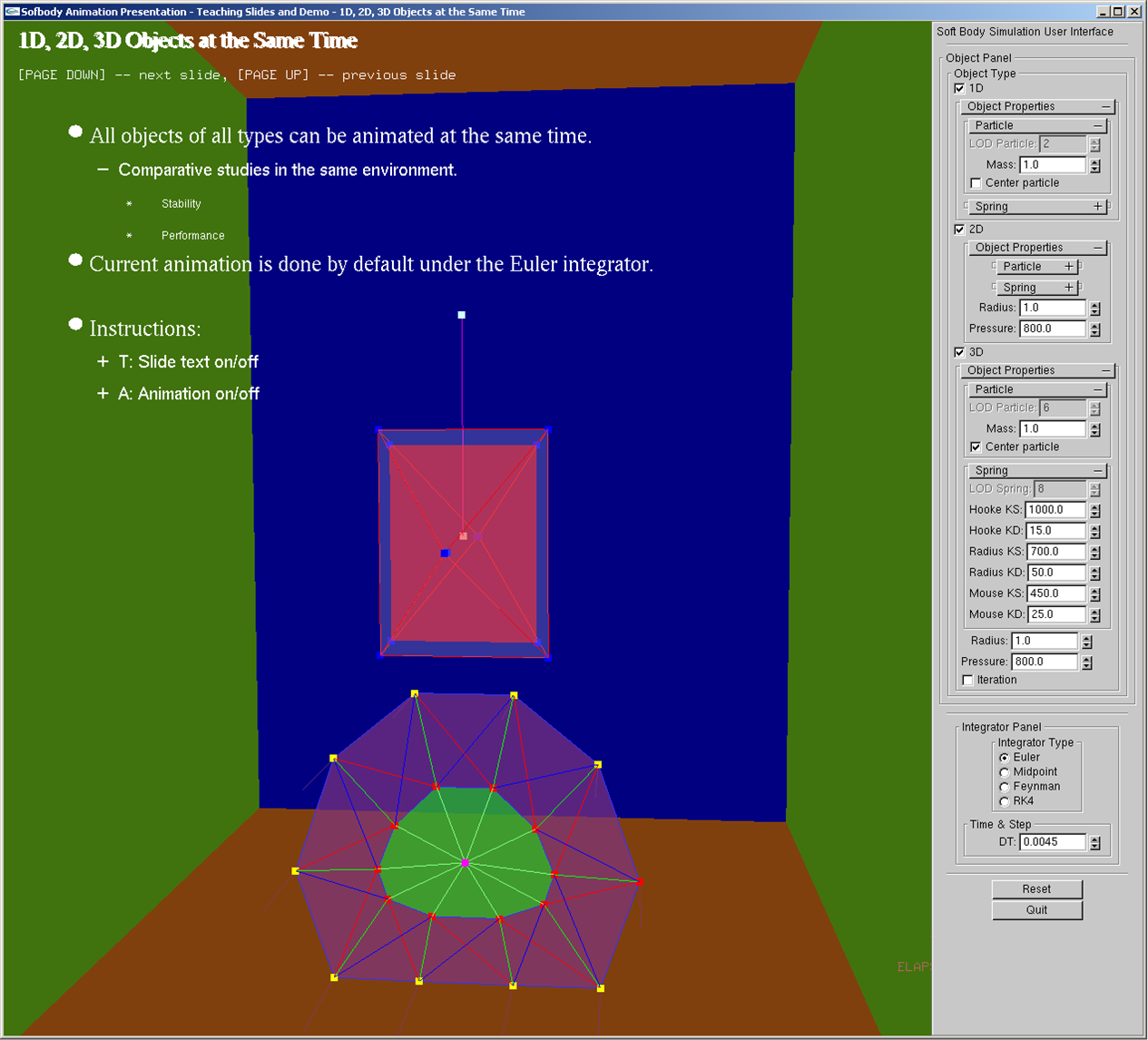}
	\caption{Three Types of Softbody Objects Simulated Together on a Single Slide Scene}
	\label{fig:Objects3}
	\end{center}
	\hrule\vskip4pt
\end{figure*}

The Softbody Simulation System's main goal is to provide real-time simulation of
a variety of softbody objects, founded in the core two- or three-
layer model for objects such as human and animal's soft parts and tissue,
and non-living soft objects, such as cloth, gel, liquid, and gas. 
Softbody simulation is a vast research topic and has a
long history in computer graphics.
The softbody system has gone through a number of iterations
in its design and development.
Initially it had limited user interface~\cite{msong-mcthesis-2007,softbody-framework-c3s2e08}.
Then the fine-grained to high-level level-of-detail (LOD) GLUI-based~\cite{gluiManual}
user interface has been added~\cite{softbody-lod-glui-cisse08},
GPU shading support was added~\cite{adv-rendering-animation-softbody-c3s2e09}
using the OpenGL Shading Language~\cite{rost2004}, a curve-based animation
was integrated, and software engineering re-design is constantly being applied.
The example of the common visual design of the LOD interactivity
interface is summarized in \xf{fig:Objects3}. The LOD components
are on the right-hand-side, expanded,
and the main simulation window is on the left (interactivity with
that window constitutes for now just the mouse drag and functional
keys).
Following the top-down approach configuration parameters, that
assume some defaults, were reflected in the GUI~\cite{softbody-opengl-slides}.

\begin{figure*}[htb!]
\hrule\vskip4pt
\begin{center}
	\subfigure[1D Elastic Object Simulation Slide]
	{\label{fig:1d-elastic-object}
	 \includegraphics[width=.47\slideimagewidth]{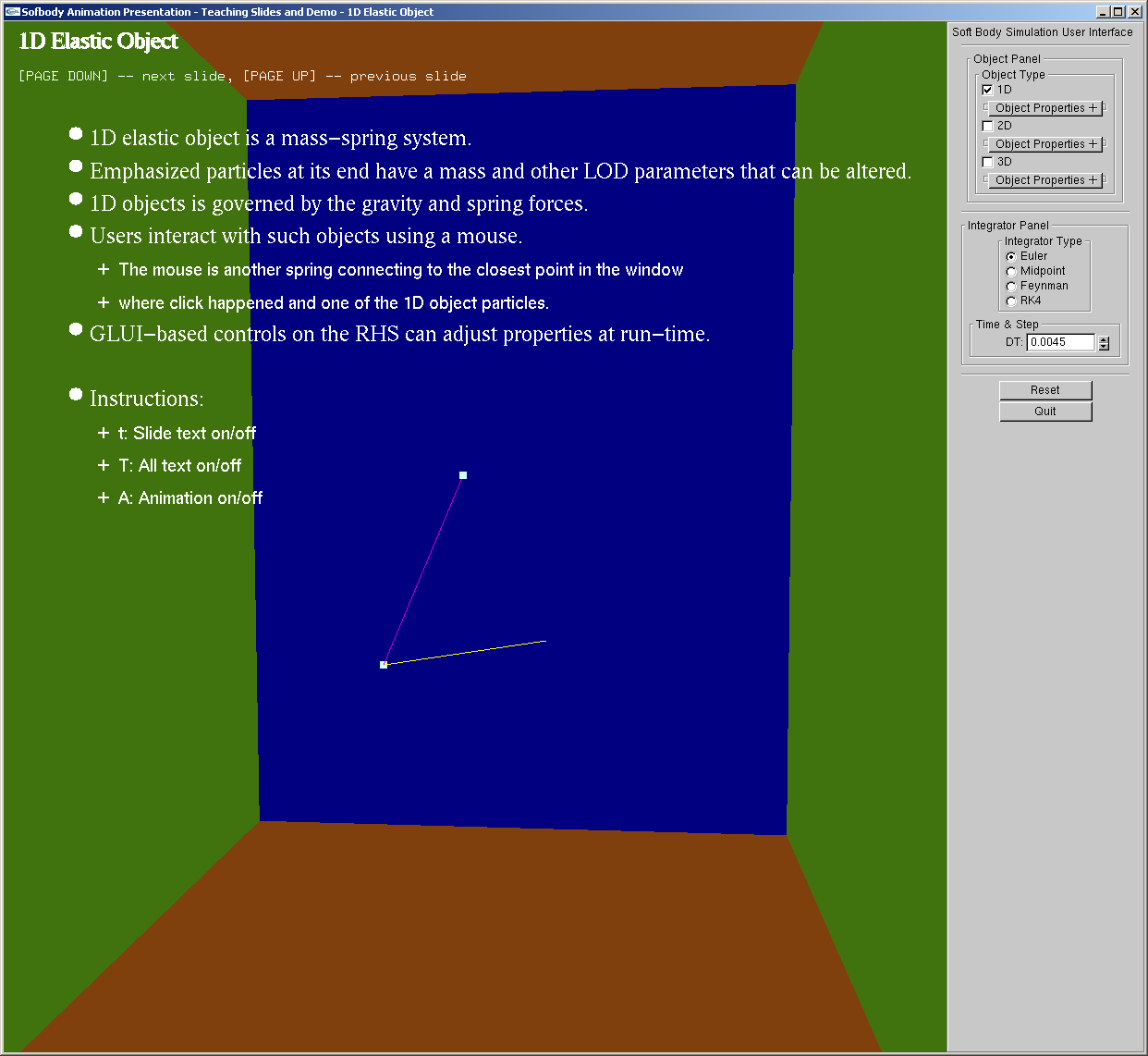}}
	\subfigure[2D Elastic Object Simulation Slide]
  {\label{fig:2d-softbody-object}
	 \includegraphics[width=.47\slideimagewidth]{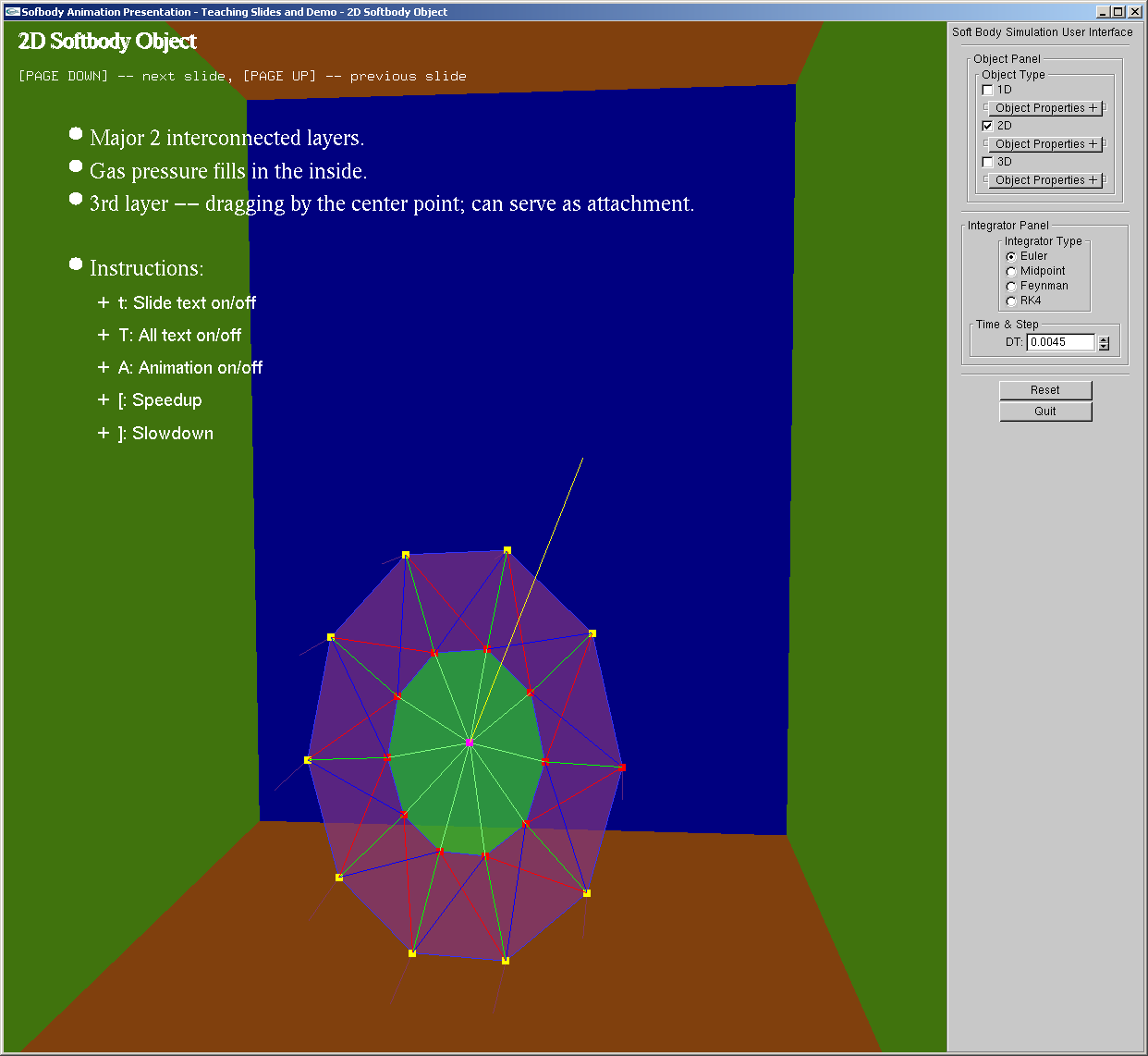}}
	\subfigure[3D Elastic Object Simulation Slide]
  {\label{fig:3d-softbody-object}
	 \includegraphics[width=.47\slideimagewidth]{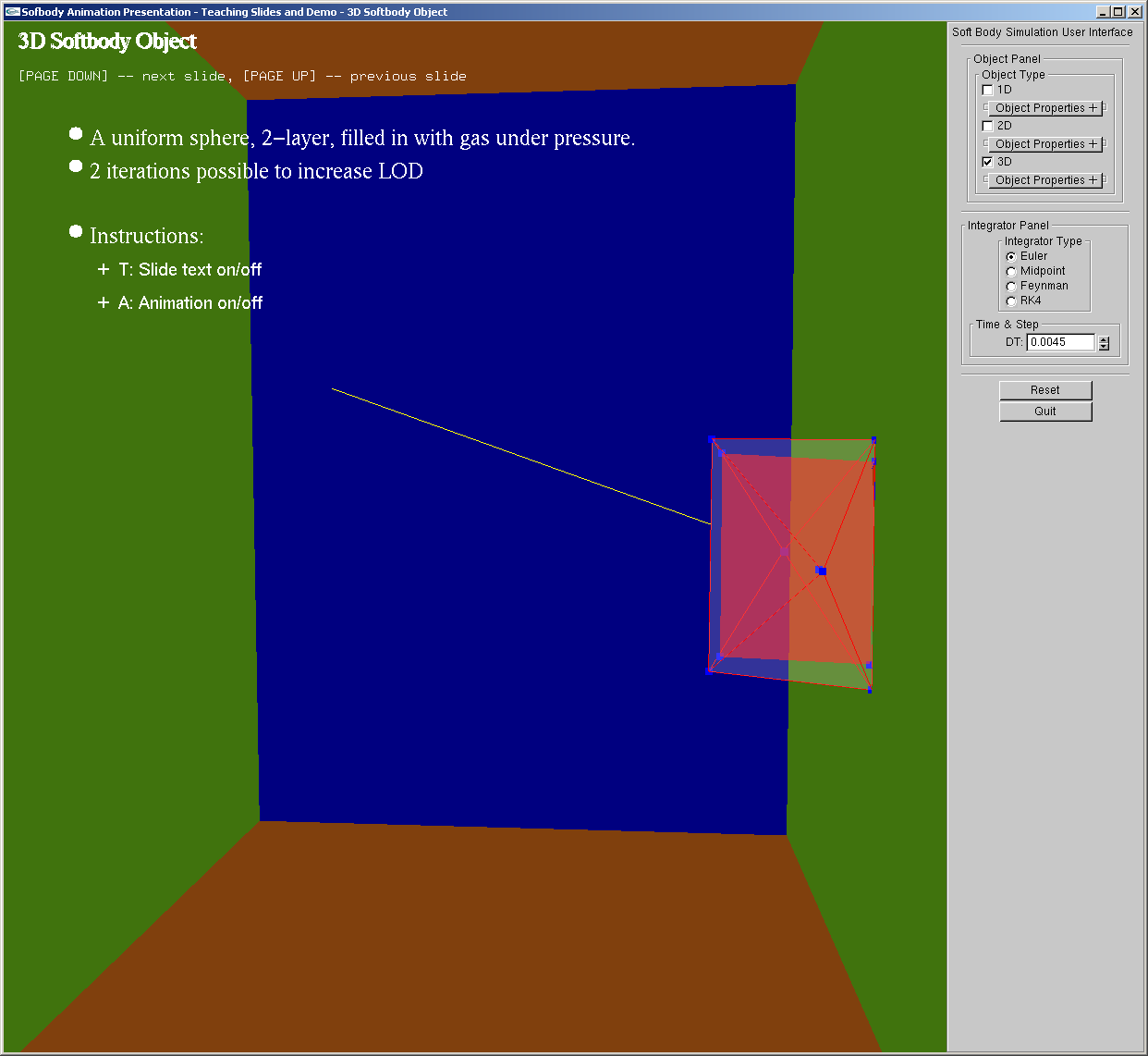}}
\caption{Simulation of Single 1D, 2D, and 3D Softbody Elastic Objects Slides}
\label{fig:object-dimensionality-slides}
\end{center}
\hrule\vskip4pt
\end{figure*}

\begin{figure*}[htb!]
\hrule\vskip4pt
\begin{center}
	\subfigure[Simulation with Euler Integrator Slide]
	{\label{fig:all-objects-euler}
	 \includegraphics[width=.47\slideimagewidth]{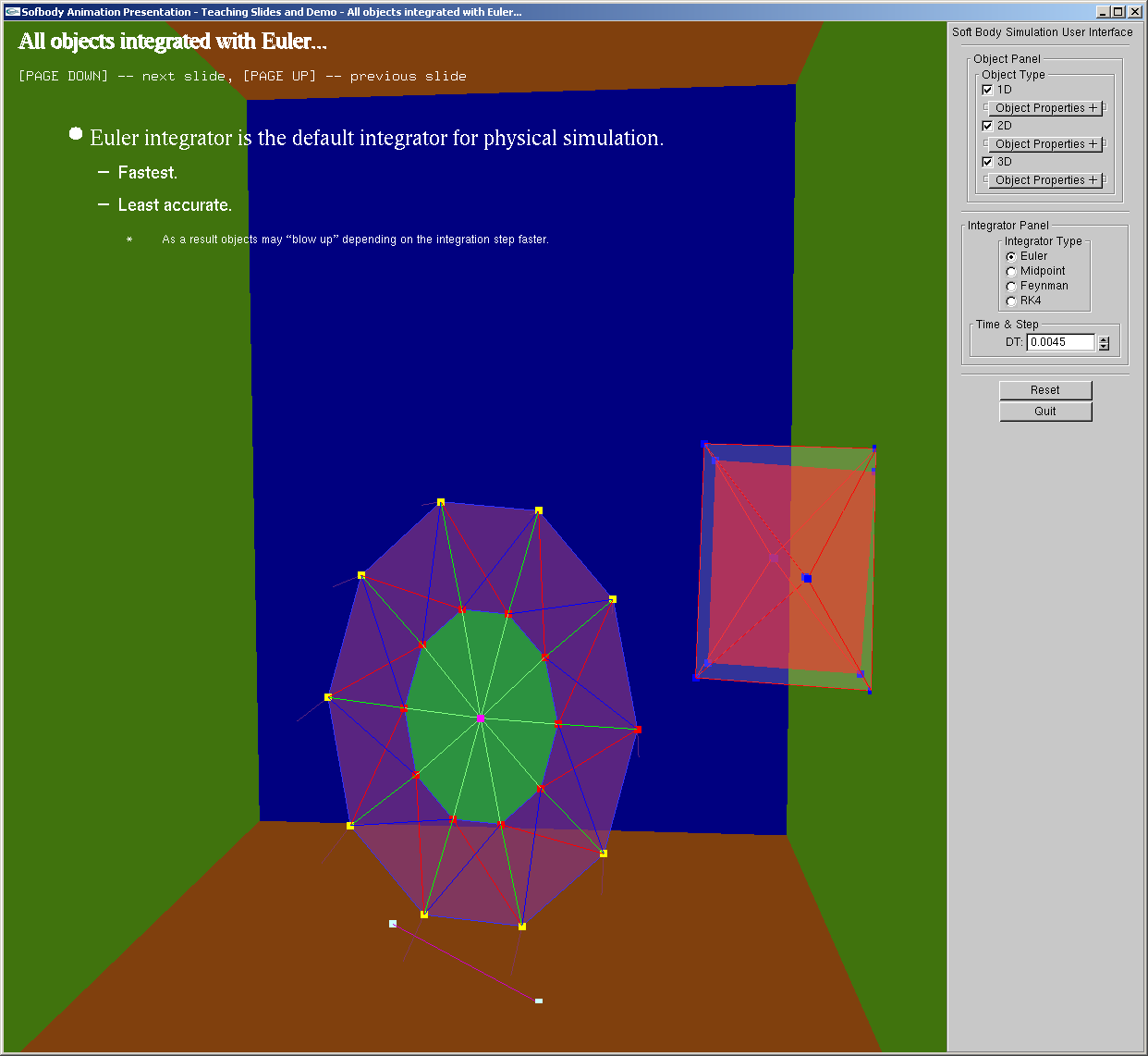}}
	\subfigure[Simulation with Midpoint Integrator Slide]
  {\label{fig:all-objects-midpoint}
	 \includegraphics[width=.47\slideimagewidth]{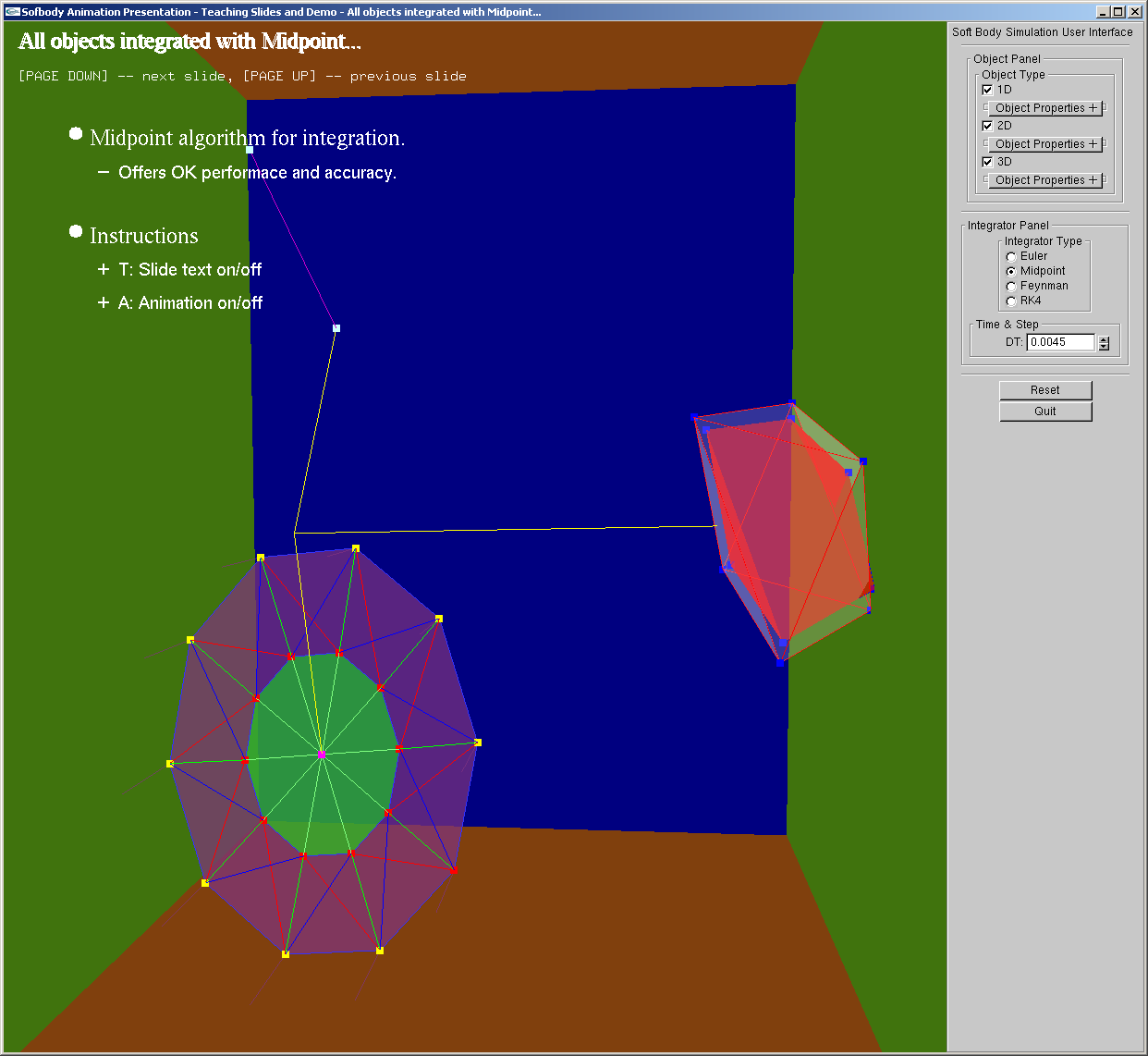}}
	\subfigure[Simulation with Feynman Integrator Slide]
  {\label{fig:all-objects-feynman}
	 \includegraphics[width=.47\slideimagewidth]{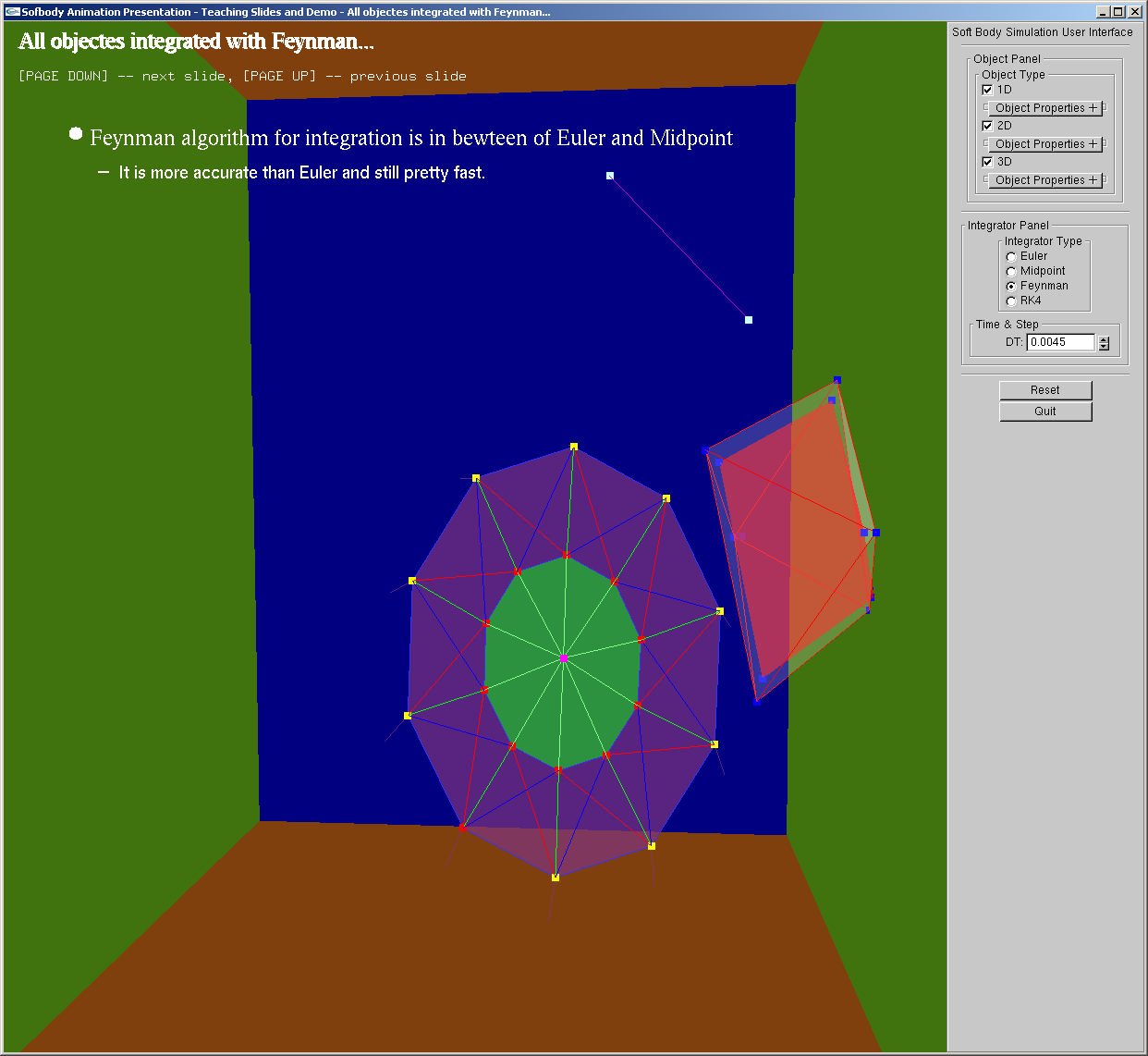}}
\caption{Simulation of all Softbody Object Types with Various Integrators Slides}
\label{fig:some-integrator-slides}
\end{center}
\hrule\vskip4pt
\end{figure*}

The core framework's design is centered around common dimensionality (1D, 2D, and 3D)
of graphical objects for simulation purposes, physics-based integrators,
and the user interaction component.
The \api{Integrator} API of the framework as of this writing is implemented by the well-known
Explicit Euler, Midpoint, Feynman, and Runge-Kutta~4 (RK4)-based integrators for their mutual
comparison of the run-time and accuracy.
The system is implemented using OpenGL~\cite{opengl,opengl-redbook}
and the {\cpp} programming language with
the object oriented programming paradigm~\cite{softbody-opengl-slides}.

This elastic object simulation system has been designed and implemented according
to the well known architectural pattern, the model-view-controller (MVC).
This pattern is ideal for real-time simulation because it simplifies
the dynamic tasks handling by separating data (the model) from user interface (the view).
Thus, the user's interaction with the software does not impact the data handling;
the data can be reorganized without changing the user interface.
The communication between the model and the view is done through the controller.
This also closely correlates to the OpenGL state machine, that is used as a core
library for the implementation~\cite{softbody-framework-c3s2e08,softbody-opengl-slides}.

\section{Methodology}
\label{sect:methodology}

\newcommand{\softbodypresentationmain}{\file{SoftBodyPresentation.cpp}}
\newcommand{\sofbodysimulationmain}{\file{SoftbodySimulation.cpp}}
\newcommand{\sofbodysimulationslide}{\api{SoftbodySimulationSlide}}
\newcommand{\apimain}{\api{main()}}
\newcommand{\apianimate}{\api{animate()}}

The methodology consists of the design and implementation
modification required for the integration followed by making
the actual presentation slides. The source code
of the presentation is a part of the learning material
along the actual content of the material presented and
is prepared as such.
Separately, both frameworks and implementing systems define
the {\apimain} function, which cannot be included into
any of the libraries (both can be compiled into the library
files to be linked into other projects) because of the linker errors
when the object code from the two or more systems is
combined into a single executable.
We therefore started a new application with a new {\apimain},
the {\softbodypresentationmain}.
Additionally, both frameworks have to declare their own
namespaces, which both have not done in the past, similarly to CUGL~\cite{cugl} because
there are some common names of variables, classes,
or functions that clash on compilation. This is an overall improvement
not only for this work, but also for any similar type
of integration with other projects (cf. \xs{sect:future-work}).
Thus, we declared the namespaces \api{softbody:} and \api{slides:}
and move the clashing variables under those namespaces.
Most of the main code from {\sofbodysimulationmain} application
is encapsulated into a generic {\sofbodysimulationslide}
class that includes the default configuration of the
softbody simulation parameters~\cite{softbody-opengl-slides}
this class is inherited by the slides that do the actual simulation
of softbody objects.
Furthermore, the concrete slides that inherit from {\sofbodysimulationslide}
are broken down into some preset distinct configuration defaults and accompanying
tidgets. They override the {\apianimate} method (the ``idle'' function) as well as
the state LOD parameters per an example slide.

For the presentation in this work the demonstration slides currently include
the following:

\begin{enumerate}
\item
\api{TitleSlide} -- a typical title slide with the lecture/presentation title and the presenter information
(see \xf{fig:welcome}).

\item
\api{TOCSlide} -- a tidget table of contents of the presentation
(see \xf{fig:contents}).

\item
\api{IntroductionSlide} -- a tidget introduction of the material
(see \xf{fig:introduction}).

\item
\api{SoftbodySimulationSlide1D} -- a slide featuring the 1D elastic
object configured by default encased in the \api{ViewBox}, see \xf{fig:1d-elastic-object}.

\item
\api{SoftbodySimulationSlide2D} -- a slide featuring the 2D softbody object configured by default,
see \xf{fig:2d-softbody-object}.

\item
\api{SoftbodySimulationSlide3D} -- a slide featuring the 3D softbody object configured by default,
see \xf{fig:3d-softbody-object}.

\item
\api{SoftbodySimulationSlideAllD} -- a slide featuring all types of softbody objects configured by default,
as shown in \xf{fig:Objects3}, included into the slide environment.

\item
\api{SoftbodySimulationSlideAllEuler} -- all three objects configured by default to animate
under the Explicit Euler integrator, see \xf{fig:all-objects-euler}.

\item
\api{SoftbodySimulationSlideAllMidpoint} -- all three objects configured by default to animate
under the Midpoint integrator, see \xf{fig:all-objects-midpoint}.

\item
\api{SoftbodySimulationSlideAllFeynman} -- all three objects configured by default to
animate under the Feynman integrator, see \xf{fig:all-objects-feynman}.

\item
\api{SoftbodySimulationSlideAllRK4} -- all three objects configured by default to animate
under the RK4 integrator, see \xf{fig:all-objects-rk4}.

\item
\api{ShortcomingsSlide} -- a slide describing the limitations of the approach
(cf. \xs{sect:limitations}, see \xf{fig:limitations}).

\item
\api{ProjectedFeaturesSlide} -- a summary of some projected features for
the future work (cf. \xs{sect:future-work}, see \xf{fig:projected-features}).

\item
\api{ConclusionSlide} -- a preliminary conclusions slide
(cf. \xs{sect:conclusion}, see \xf{fig:conclusion}).

\item
\api{ReferencesSlide} -- the list of references, see \xf{fig:references}.

\end{enumerate}

\begin{figure}[thb!]
	\hrule\vskip4pt
	\centering
	\includegraphics[width=\slideimagewidth]{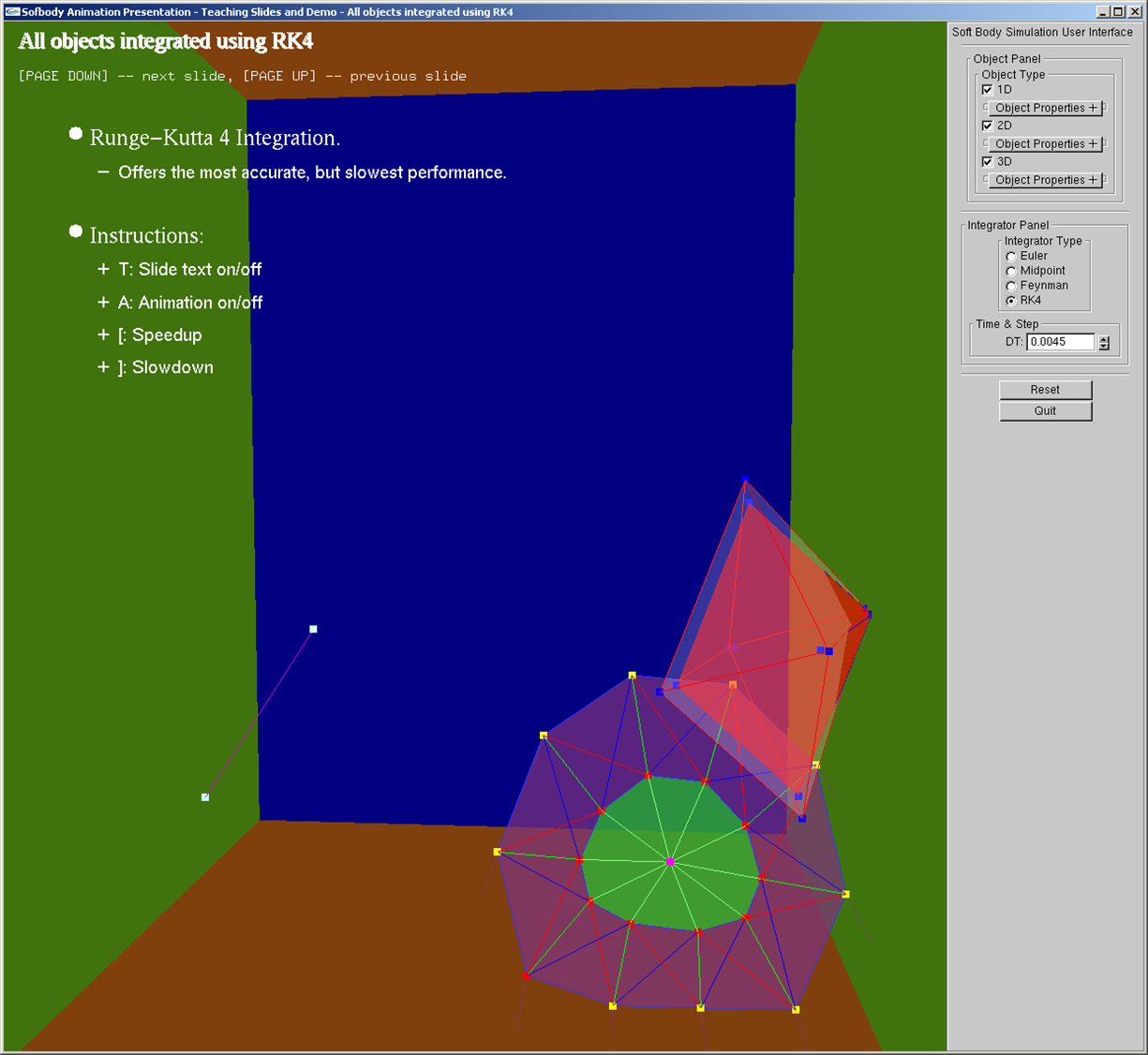}
	\caption{Simulation with Runge-Kutta 4 Integrator Slide}
	\label{fig:all-objects-rk4}
	\hrule\vskip4pt
\end{figure}

\section{Conclusions and Future Work}
\label{sect:conclusion}
\index{conclusion}

\begin{figure}[thb!]
	\hrule\vskip4pt
	\centering
	\includegraphics[width=\slideimagewidth]{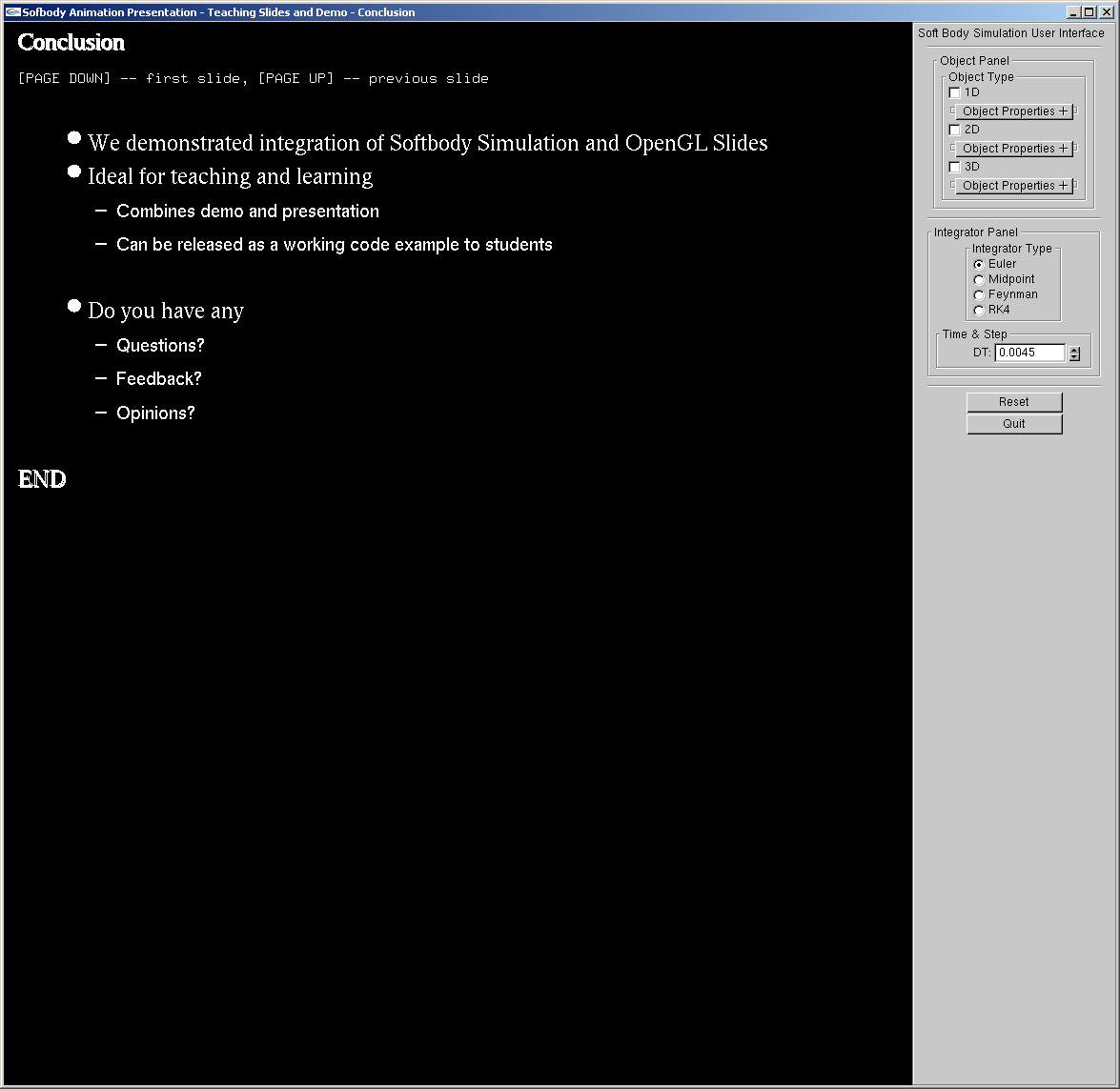}
	\caption{Conclusion OpenGL Slide Example}
	\label{fig:conclusion}
	\hrule\vskip4pt
\end{figure}

We completed the first proof-of-concept integration of
the softbody simulation system and {\oglsf} frameworks.
We made a number of slides in a OpenGL-based softbody
presentation typically found in lab/tutorial like
presentations, which are to be extended to a full
lecture-type set of slides.
This milestone significantly advances our contribution
to a good CG teaching module, suitable for use by
instructors to present the material in class as well as
for learning by providing its source code to the
students for study and extension to demonstrate their
CG projects at the end of a semester.

We have encountered some integration difficulties
due to the frameworks' original design and implementation
consideration, that we do not discuss here, but rather
discuss and generalize at length in our follow up
software engineering work in~\cite{soen-spec-cg-simulation-systems}.

We further discuss the limitation of the proposed
approach as well as the future (and ongoing) work
on these and the related projects.

\subsection{Limitations}
\label{sect:limitations}

There are some assumptions and limitations to the
approach described here; thus, it is not an all-in-one
solution, but rather for a specific purpose presentations.

\begin{itemize}
	\item 
	Assumes the CG topics taught are renderable at real-time,
	like this softbody simulation.
	\begin{itemize}
		\item 
		One can presumably also render images that were premade offline,
		but then what is the point? (Unless one is to demonstrate the
		texture mapping, etc. of course).
		\item 
		One can play AVI and HD movies in OpenGL, but again not as
		a primary learning source, though may be necessary at times
		(e.g. to teach how to play such things when/if needed).
	\end{itemize}
	\item 
	Not suitable for presentation in online conferences, so have
	to make screenshots (not a big deal, but ideally, at least
	the screenshots should be taken automatically).
	\item 
	May be hardware dependent.
	\begin{itemize}
		\item 
		Though today's commodity hardware should generally be good enough,
		but one own's laptop and the PC in a classroom may have enough
		differences to distort the presentation or render it unacceptably.
	\end{itemize}
	\item 
	Tedious to compose and debug -- need to be a programmer.
	\begin{itemize}
		\item 
		A proposal to load import info from XML or text is made.
		\item 
		Has to be done well in advance before the teaching session.
		\item
		Assumption is made the instructors and the students are able
		to do C++ or OpenGL programming in a CG programming course.
	\end{itemize}
	\item 
	Rendering is resolution-sensitive -- tidgets may have various
	spacing or stretch, not matching the softbody object being
	displayed in the scene.
	\item 
	Rendering is an issue (not supported, or tedious/difficult) for now for
	math (formulas), source code, highlighting -- can only be done as images.
\end{itemize}

\begin{figure*}[htp!]
\hrule\vskip4pt
\begin{center}
	\subfigure[Limitations OpenGL Slide Example]
	{\label{fig:limitations}
	 \includegraphics[width=.47\slideimagewidth]{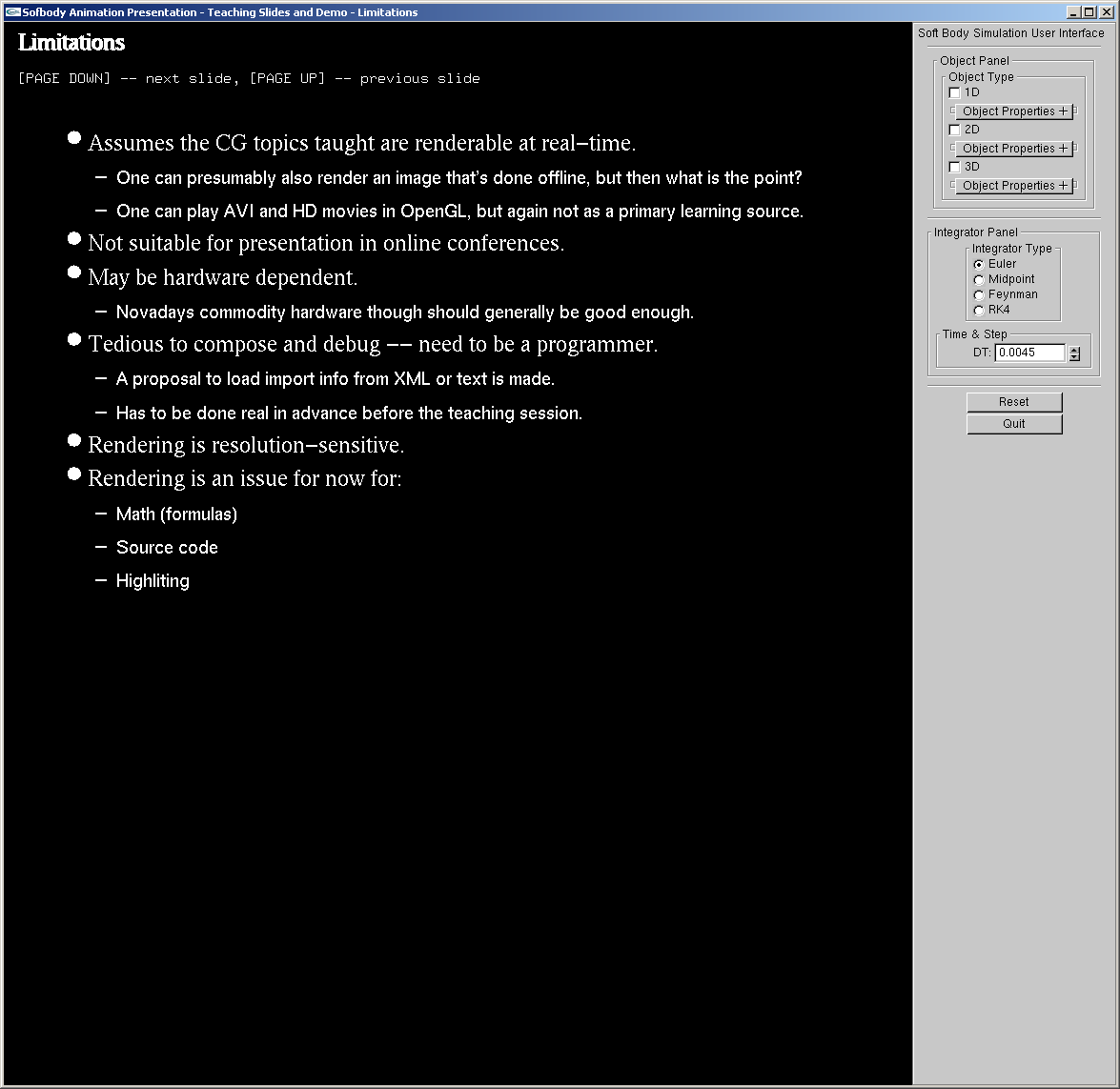}}
	\subfigure[Projected Features OpenGL Slide Example]
	{\label{fig:projected-features}
	 \includegraphics[width=.47\slideimagewidth]{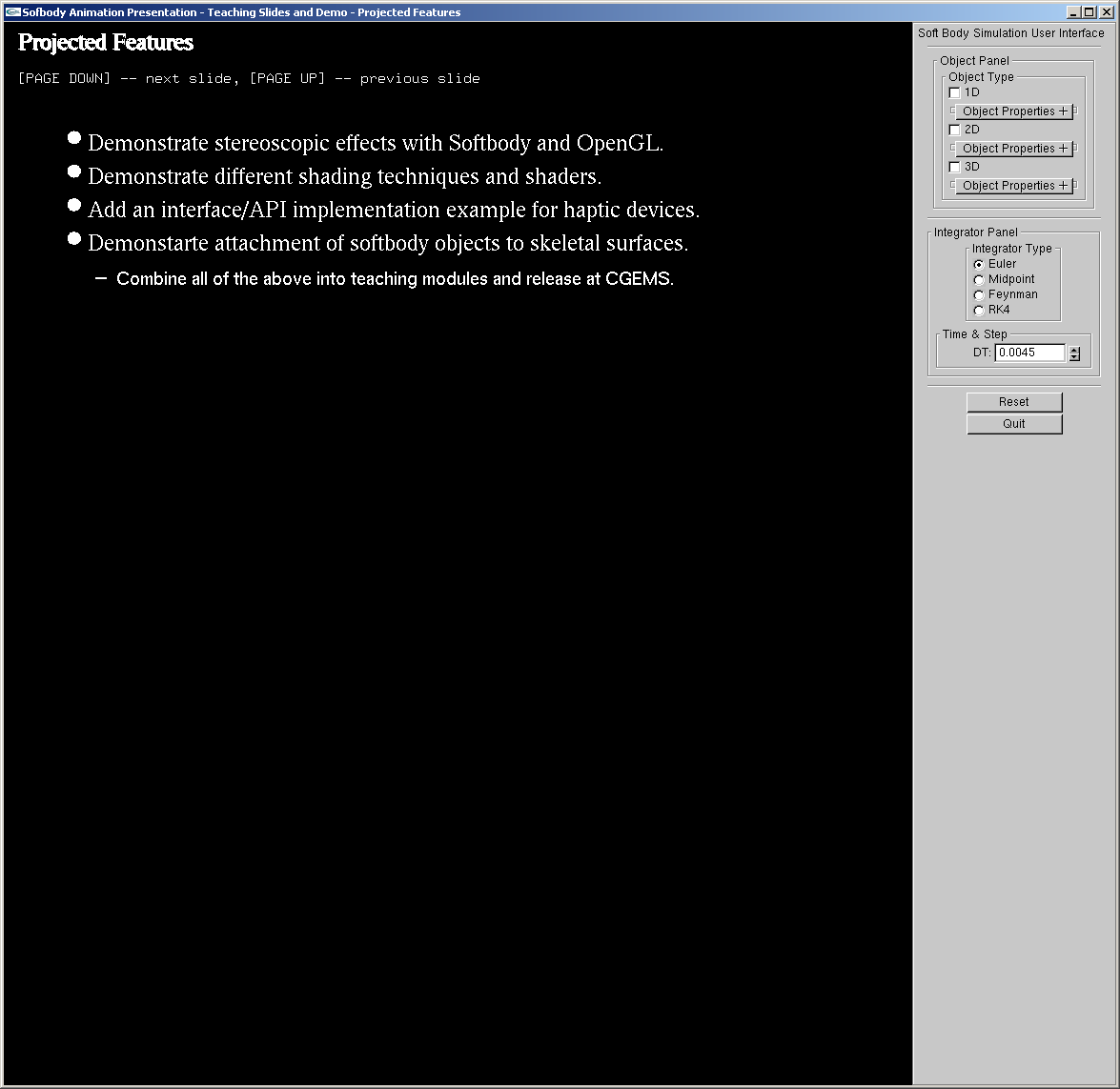}}
	\subfigure[References OpenGL Slide Example]
	{\label{fig:references}
	 \includegraphics[width=.47\slideimagewidth]{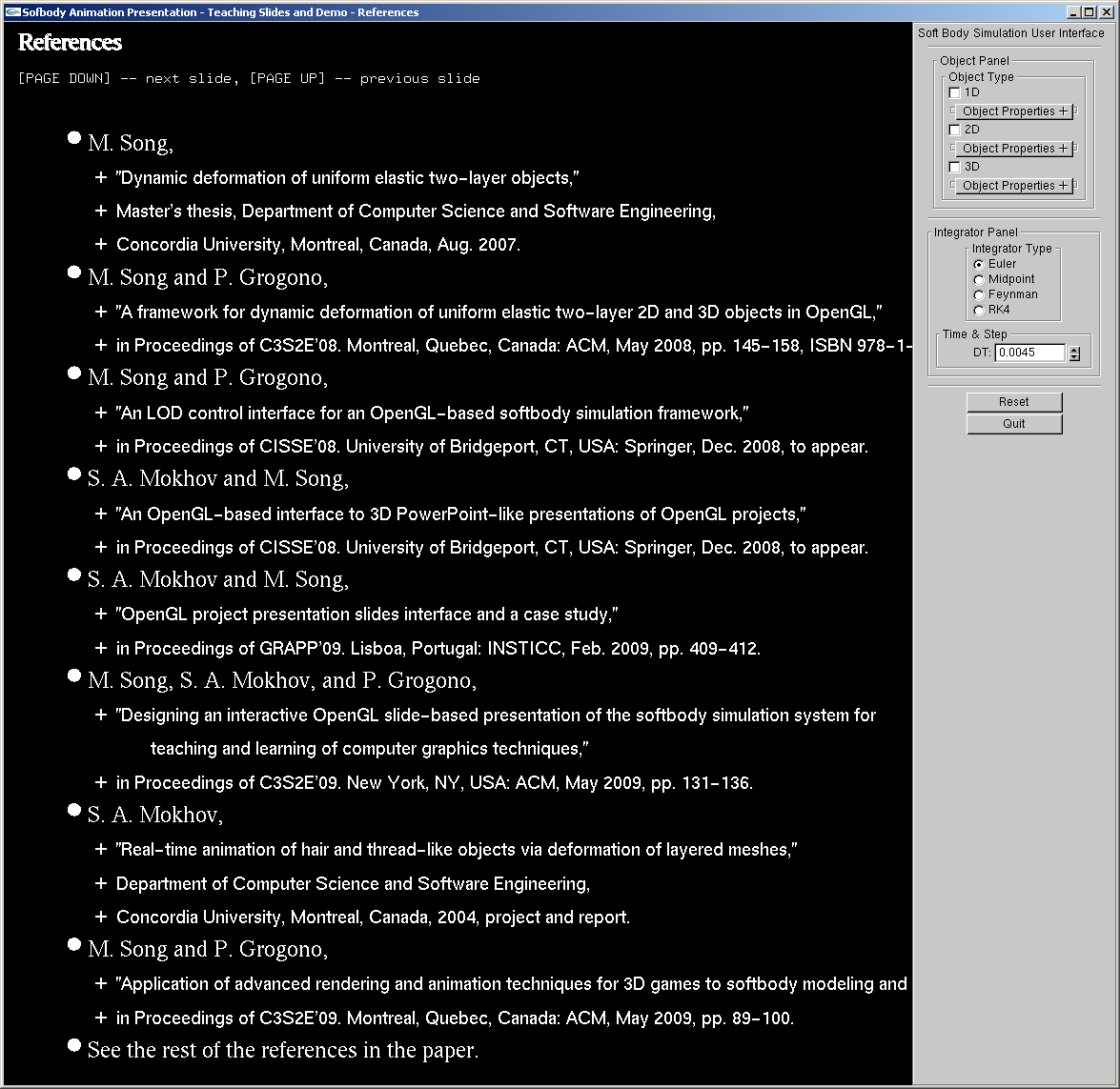}}
\caption{Concluding Slides}
\label{fig:concluding-slides}
\end{center}
\hrule\vskip4pt
\end{figure*}

\subsection{Future Work}
\label{sect:future-work}

Aside from addressing some of the limitations from the previous section,
there are a number of immediate items for the future work:

\begin{itemize}

\item
Port the source code fully to Linux and \macos{X}. Currently
it only compiles properly under \win{XP 32-bit} under Visual
Studio 2005.

\item
Release our code and documentation as open-source implementation either
a part of the Concordia University Graphics Library~\cite{cugl} and/or
as part of a Maya~\cite{maya} plug-in and as a CGEMS~\cite{cgems}
teaching module.

\item
Allow advanced interactive controls of the scenes and slides by
using haptics devices~\cite{haptics-cinema-future-grapp09} with
the force feedback, head-mounted displays and healthcare virtual reality
systems~\cite{stereo-plugin-interface}.

\item
Integrate stereoscopic effects into the presentation of softbody
objects (under way) with another open-source plug-in project under development
that implements OpenGL-based stereoscopic
effects~\cite{stereo3d,arloader-thesis-08,stereo-plugin-interface}.

\item
While working on this and other integration efforts, take
down and formalize software engineering requirements for systems
like ours to simplify the future development and integration
process of academic and open-source OpenGL and CG frameworks
and systems for physical based animation and beyond~\cite{soen-spec-cg-simulation-systems} (in progress).

\item
Provide automatic loading and display of the softbody simulation source code
on the slides with breakdown onto multi-page slides.

\item
Demonstrate different softbody shading techniques and shaders via the OpenGL slides.
We already implemented the first draft version
of a vendor-independent API for shader use within the softbody system
and provided two implementations of that API -- one that loads GLSL
vertex and fragment shaders and the other that loads
the cross-vendor assembly language for shaders.

\item
Demonstrate attachment of softbody objects to skeletal surfaces via the slides.

\end{itemize}

\section*{Acknowledgments}

This work is partially funded by NSERC, FQRSC, and the Faculty of Engineering
and Computer Science, Concordia University, Montreal, Canada.


\bibliographystyle{plain}
\bibliography{softbody-opengl-slides-arXiv}

\end{document}